\documentclass[prb,amsmath,amssymb,twocolumn]{revtex4-2}
\usepackage[utf8]{inputenc}
\usepackage{amsmath, amsfonts}
\usepackage{amscd}
\usepackage{amsthm}
\usepackage{amssymb}
\usepackage{float}
\usepackage{cases}
\usepackage{mathtools}
\usepackage{graphicx}
\usepackage{physics}
\usepackage{natbib}
\usepackage{xcolor}
\usepackage{bbm}
\usepackage{booktabs}
\usepackage{makecell}
    {\clearpage\pagebreak[4]\global\pdfpageattr\expandafter{\the\pdfpageattr/Rotate 90}}%
    {\clearpage\pagebreak[4]\global\pdfpageattr\expandafter{\the\pdfpageattr/Rotate 0}}%

\begin{document}

\newpage

\title{Disease Gene Prioritization With Quantum Walks}

\author{Harto Saarinen}
\thanks{These two authors contributed equally.}
\affiliation{Algorithmiq Ltd, Kanavakatu 3 C, FI-00160 Helsinki, Finland}
\affiliation{Complex Systems Research Group, Department of Mathematics and Statistics, University of Turku, FI - 20014 Turun Yliopisto, Finland}

\author{Mark Goldsmith}
\thanks{These two authors contributed equally.}
\affiliation{Algorithmiq Ltd, Kanavakatu 3 C, FI-00160 Helsinki, Finland}
\affiliation{Complex Systems Research Group, Department of Mathematics and Statistics, University of Turku, FI - 20014 Turun Yliopisto, Finland}

\author{Rui-Sheng Wang}
\affiliation{Department of Medicine, Brigham and Women's Hospital, Harvard Medical, Boston, MA 02115}

\author{Joseph Loscalzo}
\affiliation{Department of Medicine, Brigham and Women's Hospital, Harvard Medical, Boston, MA 02115}

\author{Sabrina Maniscalco}

\affiliation{Algorithmiq Ltd, Kanavakatu 3 C, FI-00160 Helsinki, Finland}

\date{\today}

\begin{abstract}
Disease gene prioritization assigns scores to genes or proteins according to their likely relevance for a given disease based on a provided set of seed genes.
Here, we describe a new algorithm for disease gene prioritization based on continuous-time quantum walks using the adjacency matrix of a protein-protein interaction (PPI) network. Our algorithm can be seen as a quantum version of a previous method known as the diffusion kernel, but, importantly, has higher performance in predicting disease genes, and also permits the encoding of seed node self-loops into the underlying Hamiltonian, which offers yet another boost in performance.
We demonstrate the success of our proposed method by comparing it to several well-known gene prioritization methods on three disease sets, across seven different PPI networks. In order to compare these methods, we use cross-validation and examine the mean reciprocal ranks and recall values. We further validate our method by performing an enrichment analysis of the predicted genes for coronary artery disease. We also investigate the impact of adding self-loops to the seeds, and argue that they allow the quantum walker to remain more local to low-degree seed nodes.
\end{abstract}

\maketitle

\section{Introduction}

The utilization of network modelling has proven to be an effective technique for studying the structure and dynamics of biological systems \cite{Yan2017, Stelzl2006}. Consequently, there has been an increasing effort through biophysical and high-throughput methods to form protein-protein interaction (PPI) networks that consist of the physical and/or functional interactions between human proteins. This type of complex network, called the human interactome, sets the basis for the field of network medicine \cite{Barabasi2011, Loscalzo2011, Lee2019}. 

One of the main propositions of network medicine is that a disease phenotype is rarely a consequence of abnormal effects in a product of a single gene, but, rather, the effects are scattered across multiple gene products interacting in the human interactome \cite{Goh2007, Menche2015}. These interacting proteins associated with a given disease thus form a subnetwork and are then expected to gather in a local neighbourhood in the human interactome \cite{Barabasi2011}. As the proteins in these disease subnetworks are collectively involved in the development or progression of a disease, they offer key insights into the underlying molecular mechanisms and biological processes driving the disease. However, understanding these molecular disease mechanisms can be time-consuming and require significant resources when using high-quality and/or small-scale experiments. Thus, locating these disease neighbourhoods, called disease modules \cite{Loscalzo2011}, has been a major challenge in the field with a clear need for improved computational methods. 

In the context of disease module identification, the typical scenario involves a predefined set of proteins known as seed proteins, which have been carefully curated and experimentally validated for their association with a specific disease. However, these seed proteins often form an incomplete subnetwork that does not fully represent the expected disease module. The primary objective is to predict the entire disease module by leveraging the structure of the PPI network and the seed proteins. This problem is commonly referred to as disease gene prioritization, as the aim is to systematically incorporate additional proteins into the module based on their likelihood of being disease-associated. Once the seed genes are identified and mapped to the PPI, network-based connector proteins provide the missing links needed to define the disease module. Thus, by employing computational methods and network analysis techniques, researchers can enhance the comprehensiveness of the disease module and identify potential candidate genes for further experimental investigation. 

Because the network approach to diseases has already demonstrated its effectiveness for multiple diseases \cite{Sharma2015, Wang2021, Petro2013, Pandey2023, Pankaj2023, Lee2020}, it is of utmost importance to develop and elaborate on methods that identify disease modules. Unfortunately, there are a few well known challenges in identifying the disease modules. On the one hand, the PPI networks tend to be very incomplete, with various estimates suggesting that they account for approximately 20-30\% of the total connections within the interactome \cite{Hart2006, Vidal2011, Venkatesan2009, Luck2020}. Hence, the structure of the human interactome is partly unknown and the disease modules tend to be more scattered around the interactome than is expected. On the other hand, as noted in \cite{Ghiassian2015diamond}, the state of the art community detection algorithms, which are shown to work well in other network clustering tasks, tend to perform very poorly in locating these disease modules. Hence, there has been increasing effort applied to the development of methods that are specifically designed to infer the disease modules in these very incomplete networks.

We introduce a novel method centred around quantum walks on the interactome. Continuous-time quantum walks, initially proposed in~\cite{Farhi1998}, are the quantum analogues of continuous-time classical random walks, which describe the propagation of a particle over a graph. Together with their discrete-time counterpart~\cite{Aharonov1993}, they have received much attention for their applications in quantum information processing~\cite{Kempe2003, Venegas2012}, quantum computation~\cite{Childs2009}, and quantum transport~\cite{Mulken2011}. While the methods that we describe here are \textit{{quantum-inspired}}, since they are implemented classically, we can foresee that these algorithms will be even more efficient if run on quantum devices. Continuous-time quantum walks have already been implemented on various physical platforms~\cite{Manouchehri2014}, including optical setups~\cite{Young2022, Wang2020, Tang2018, Peruzzo2010, Preiss2015} and superconducting devices~\cite{Gong2021, Yan2019}, and they can also be simulated on gate-based quantum computers~\cite{Loke2017, Qiang2017}.

In general, random walk methods are known to perform well in a variety of tasks \cite{Xia2020}, and have also been used for disease module detection \cite{Petro2013, Kohler2008, Joodaki2021, Xie2012, Li2010, Gentili2022}. However, most of these methods are based on discrete-time random walks on the network or its modifications, while their continuous-time counterparts have not been studied as extensively, even though they seem to have rather competitive performance \cite{Kohler2008}. Quantum walks have not been previously used in disease module identification. Based on these observations, we propose a new disease gene prioritization method based on continuous-time quantum walks using the PPI adjacency matrix (QA). The choice of continuous-time quantum walks is two-fold. Firstly, quantum walks have been shown to perform competitively in other network applications such as link prediction \cite{Qian2017, Omar2023, Goldsmith2023} and spatial search \cite{Malmi2022}. Secondly, quantum walks can work analogously to continuous-time classical random walk methods (such as the diffusion kernel \cite{Kohler2008}, described later) but offer more flexibility in terms of the dynamics that they can produce, which allows them to be modified suitably for the disease module identification task.

In the disease module identification problem, evaluating the performance of different methods is not straightforward since the ground truth of the predicted modules is unknown. In this study, we compare the performance of our proposed method against a disease module detection algorithm (DIAMOnD) \cite{Ghiassian2015diamond}, neighbourhood scoring (NBR) \cite{Navlakha2010}, random walk with restart (RWR) \cite{Kohler2008}, and diffusion kernel (DK) \cite{Kohler2008} methods, using cross-validation similar to standard practices in machine learning and the link prediction literature (see, e.g., \cite{Hastie2009, Lu2011survey} and references therein). We evaluate the methods on seven PPI networks from various sources and three different data sets from different databases for disease seeds. However, to compare the methods in this manner, they must be able to yield disease modules of any size. Many algorithms such as SCA \cite{Wang2018sca}, TOPAS \cite{Buzzao2022topas}, and DOMINO \cite{Levi2021domino} use Steiner trees or other ways of connecting the seed proteins so that the size of the predicted module varies greatly from disease to disease. Importantly, it is not a hyperparameter that the user can control. Thus, our cross-validation-based comparison is not feasible for these methods. In addition, all of those methods aim to form a single necessarily connected module for all diseases, which might ignore crucial disease components \cite{Agrawal2018}.

\section{Material and methods}

\subsection{The setup}

Consider a protein-protein interaction network modelled by an undirected graph $G=(V,E)$, where $V$ is the set of proteins (nodes) of size $n$ and $E$ is the set of interactions (edges). The \emph{adjacency matrix} of $G$ is the $n \times n$ matrix defined by 
\begin{equation*}
    A = (A_{ij}) =
    \begin{cases}
    1, & \text{ if } (i,j) \in E, \\
    0, & \text{ if } (i,j) \not \in E.
    \end{cases}
\end{equation*}
The \emph{network Laplacian} is defined as $L=D-A$, where $D$ is the diagonal \emph{degree matrix} given by $D = \text{diag} \left(\sum_{j}A_{1j}, \ldots, \sum_{j}A_{nj}\right)$.

A \emph{disease module} $DM$ in the network $G$ is a (connected) subnetwork of $G$ that contains proteins $S = (s_1, \ldots, s_d)$ called \emph{seed proteins}. The seed proteins should be understood as a set of proteins that by definition are part of the disease module while the rest of the module $DM \setminus S$ is unknown. Thus, the problem of locating the disease module $DM$ of unknown size is to find the proteins in $G$ associated with a disease given a set of seed proteins $S$.

\subsection{Continuous-time quantum walks}

In the classical continuous-time random walk on a network every edge of the network is associated with an independent Poisson process with unit intensity. When the walker is at some node, it will remain there until one of the Poisson processes at a neighbouring edge jumps, at which point the walker follows that edge to the corresponding neighbour, and then the process repeats. Working out the mathematical details leads to a rather simple closed-form formula for the evolution of the walker.

In contrast to a classical random walk, a quantum walk on a network evolves according to the laws of quantum physics and its evolution is governed by the Schrödinger equation. Consequently, the paths of the walker across the network can interfere constructively or destructively. This interference can cause the evolution of the quantum walker to be significantly different from the classical one \cite{Aharonov1993, Childs2002}.

A continuous-time quantum walk \cite{Farhi1998} on a graph $G$ is defined by considering the Hilbert space $\mathcal{H}$ spanned by the orthonormal vectors $\{\ket{i}\}_{i=1}^n$, corresponding to the nodes of the network, and the unitary transformation $e^{-itH},$ where $H$ is the Hamiltonian that is based on the structure of the network under consideration. Using this unitary transformation, the initial state vector $|\psi(0)\rangle$ in $\mathcal{H}$ evolves via
\begin{equation}\label{eq:QWstatevector}
\left| \psi (t)\right\rangle = e^{-itH} \left| \psi (0)\right\rangle.
\end{equation}
In general, the Hamiltonian $H$ can be any Hermitian matrix related to $G$ as long as it describes the structure of the network \cite{Venegas2012}, but usually the Laplacian $L$ or the network adjacency matrix $A$ is used \cite{Wong2016}. This is in contrast to the classical case, where the Laplacian must be used, giving the quantum walk more flexibility in terms of the dynamics. In this paper we exploit this property by modifying the adjacency matrix by adding a constant real number $\alpha$ to the diagonals corresponding to the given seed proteins. We note that this is equivalent to adding $\alpha$ self-edges at the seed proteins in the network and, consequently, it increases the likelihood of the walker remaining in the vicinity of the seed nodes for a longer period of time (see the ablation study in the Discussion section for details). This effect is very similar to lazy classical random walks. Thus, as the Hamiltonian we use
\begin{equation}
\label{eq:seed_diag}
    A_S = A + \alpha \, \text{diag} (v_S),
\end{equation}
where $v_S$ is a binary vector defined by $v_i = \mathbbm{1}_{\{ i \in S \}}$, where $S$ is the set of seed proteins and $\mathbbm{1}$ is the indicator function.

In order to obtain a probability transition matrix from the Hamiltonian, we evolve the system for a time $t$ and perform a measurement, which can be done by taking the square of the modulus of the entries of the unitary operator $e^{-it A_S},$ where $i=\sqrt{-1}.$  The entries of the probability transition matrix are
\begin{equation}
\label{eq:QRWprobabilitymatrix}
    P_{uv} (t)= |\langle v | e^{-it A_S} | u \rangle|^2.
\end{equation}
Note that, contrary to the classical case, where randomness comes from stochastic transitions between states, state transitions are deterministic in the quantum walk, with randomness resulting from the measurement and collapse of the wave function. 

Once these transition probabilities are calculated, we proceed similarly to the diffusion kernel method initially proposed in \cite{Kohler2008}, which postulates that a protein is more likely to be associated with a disease if the walker is likely to transition from that protein to any of the seed proteins. Thus the likelihood score $L_t(v)$ for protein $v$ is computed by summing the probabilities for the walker to move from $v$ to any node in the seed set $S$, computed at time $t$. More specifically,
\begin{equation*}
    L_t(v) = \sum_{s \in S} P_{vs}(t).
\end{equation*}
In this case, $t$ is a hyperparameter that can be chosen for the data set in question.

When considering a specific disease in the disease module identification task, we do not need the whole matrix exponential, but, rather, its action on the seed vector $v_S$. Consequently, calculating the scores for all the considered networks can be an efficient process \cite{al2011computing, higham2010computing}.

\subsection{Data}
\subsubsection{Human Interactome Networks}
We tested our methods on variety of different human PPI networks, which have previously been used for disease module detection. The \textit{GMB} PPI was constructed from seven different sources, described in \cite{Menche2015}; the \textit{WL} PPI integrated data from protein-protein interactions, protein complexes, kinase-substrate interactions, and signalling pathways \cite{Wang2021network}; and the 5 PPI networks \textit{BioGRID}, \textit{STRING}, \textit{APID}, \textit{HPRD}, and \textit{IID} were retrieved from well-known PPI databases and made available in \cite{Lazareva2021limits}.

\begin{figure}[!htb] 
 \includegraphics[width=0.95\linewidth]{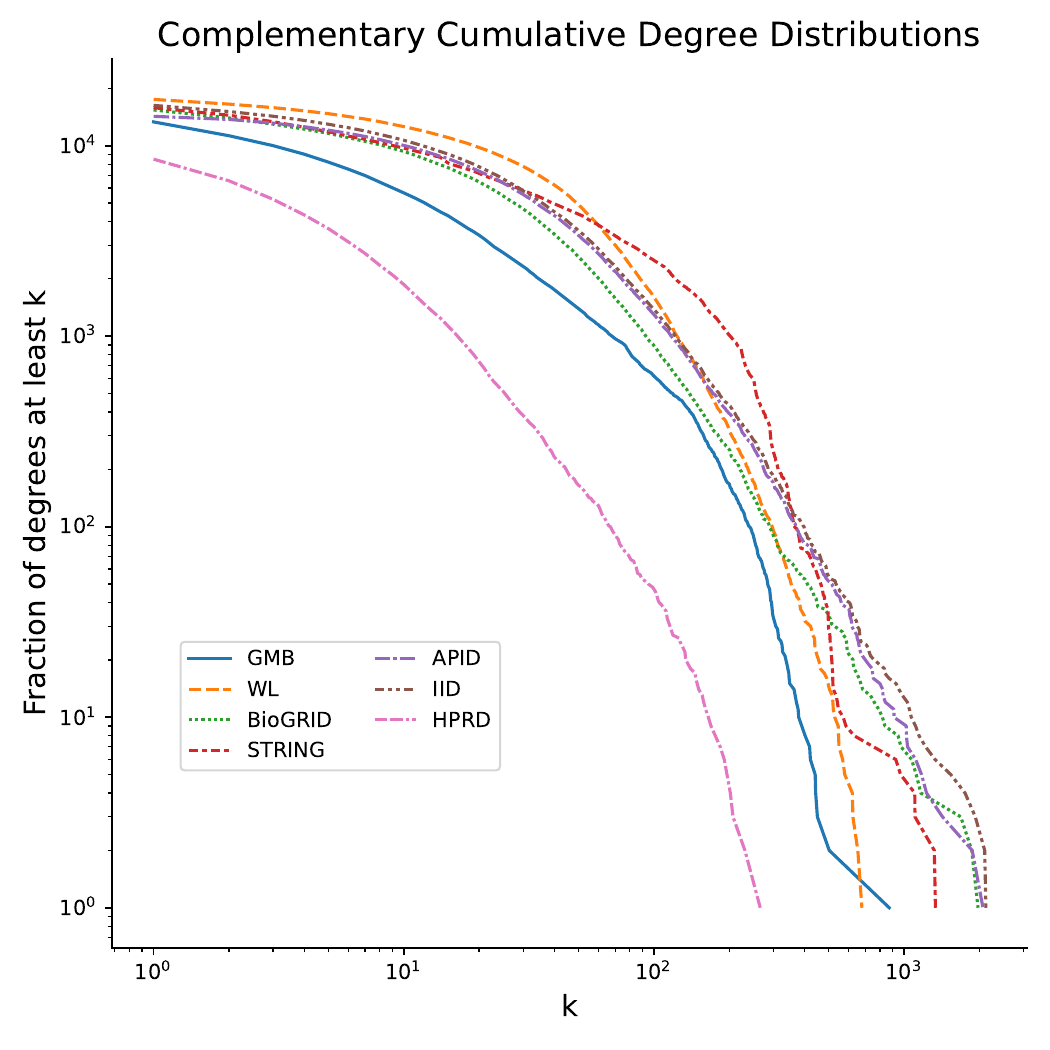}
 \caption{Complementary cumulative degree distributions. For each degree value $k$ ($x$-axis), the number of nodes with degree greater than or equal to $k$ ($y$-axis) is shown, each on a logarithmic~scale.}
 \label{fig:degdist}
\end{figure}

Some statistics of these networks are listed below in Table~\ref{tab:network_properties}, and their degree distributions are shown in Figure \ref{fig:degdist}. We observe from these statistics that the networks have high clustering and that they are very sparse. 
Furthermore, the networks are approximately scale-free~\cite{Barabasi1999}, which is typical of biological networks. One distinguishing feature of PPI networks compared to most other complex networks is that they may sometimes contain self-edges, which represent the ability of a protein to interact with itself.

\begin{table}[!htb]
    \scalebox{0.95}[1.0]{\begin{tabular}{|c|c|c|c|c|c|c|c|}
    \hline
    \textbf{Network} & \boldmath{$|V|$} & \boldmath{$|E|$} & \boldmath{$\langle k \rangle$} & \boldmath{$\rho$} & \boldmath{$C$} & \boldmath{$A$} & \textbf{SIPs} \\
\hline
HPRD    &   8498 &   33935 &        7.987 &    0.001 &            0.109 &         -0.034 &           0 \\
GMB     &  13329 &  141150 &       21.179 &    0.002 &            0.174 &          0.115 &        2794 \\
APID    &  14257 &  292964 &       41.098 &    0.003 &            0.122 &         -0.046 &           7 \\
BioGRID &  15400 &  237045 &       30.785 &    0.002 &            0.104 &         -0.063 &           2 \\
STRING  &  15821 &  387175 &       48.944 &    0.003 &            0.407 &          0.182 &           7 \\
IID     &  16280 &  314956 &       38.692 &    0.002 &            0.116 &         -0.065 &        4063 \\
WL      &  17491 &  354640 &       40.551 &    0.002 &            0.082 &         -0.034 &           0 \\ \hline
    \end{tabular}}
    \caption{Some properties of the networks that were tested. $|V|:$ number of nodes, $|E|:$ number of edges, $\langle k \rangle :$ average degree, $\rho:$ network density, $C:$ average clustering coefficient, $A:$ assortativity, $\text{SIPs}$: number of self-interacting proteins (self-edges).}
    \label{tab:network_properties}
\end{table} 
\subsubsection{Disease Genes}
We gathered disease data from three main sources: Open Targets (OT) \cite{Ochoa2022opentargets}, DisGeNET (DGN) \cite{Pinero2019disgenet}, and the disease data provided in \cite{Ghiassian2015diamond} (GMB). Open Targets and DisGeNET are large-scale databases that integrate data from a combination of different sources such as GWAS databases, genetics, drugs, animal models, and the scientific literature. The GMB data set from \cite{Ghiassian2015diamond} was curated by experts from OMIM \cite{Hamosh2002,Mottaz2008} and the PheGenI database \cite{Ramos2014}. 

The Open Targets and DisGeNET sources each include thousands of diseases, while the GMB data set from \cite{Ghiassian2015diamond} only contains 70 expert selected diseases. In order to use more manageable disease sets from Open Targets and DisGeNET, we filtered the disease sets using a ranking (score) of the disease gene associations provided by these data sets. For Open Targets, we only used disease-gene associations with a score of at least 0.6; for DisGeNET we only used disease-gene associations with a score of at least 0.3 (so that there is likely at least one `curated` source), and ensured that seed genes have a disease specificity index of at least 0.5. Finally, for each PPI, we only used the diseases whose PPI coverage contains at least 15 genes after the above filtering. Table \ref{Table:Diseases} shows the number of diseases used from each data set, on each PPI considered. 

\begin{table}[!htb]
\begin{tabular}{lccccccc}
\toprule
Network &  APID &  BioGRID &  GMB &  HPRD &  IID &  STRING &   WL \\
Disease set &       &          &      &       &      &         &      \\
\midrule
DGN         &   358 &      354 &  333 &   263 &  380 &     379 &  379 \\
GMB         &    64 &       63 &   65 &    58 &   64 &      63 &   64 \\
OT          &    49 &       49 &   48 &    31 &   50 &      50 &   49 \\
\bottomrule
\end{tabular}
\caption{Number of diseases for each disease set and network.}
    \label{Table:Diseases}
\end{table}

\subsection{Related works}

In order to assess the performance of our method, we selected four other disease gene prioritization methods previously considered in the literature for comparison: diffusion kernel (DK) \cite{Kohler2008},
random walk with restart (RWR) \cite{Kohler2008}, DIAMOnD (Dia) \cite{Ghiassian2015diamond}, and neighbourhood scoring (NEI) \cite{Navlakha2010}. These methods are briefly described here. \\

\noindent
\textbf{Diffusion kernel (DK). \cite{Kohler2008}}
A continuous-time classical random walk on a network is a Markov process with state space $V$ characterized by a rate matrix $L$ and initial distribution $\mathbf{p}(0)$ over a set of nodes. Hence, the dynamics are governed by the \emph{diffusion kernel} 
\begin{equation}\label{eq:RWprobabilitymatrix}
    \mathbf{p}(t) = \mathbf{p}(0)e^{-tL}.
\end{equation}
These transition probabilities are then used to compute scores for proteins that are not in the seed set by calculating
\begin{equation*}
    S(i) = \sum_{s \in S} P_{is}.
\end{equation*}
This model has a single hyperparameter, $t$. For our experiments, we used $t=0.3$ since this value provided the strongest results on the GMB network and data set. No discussion of the setting of this hyperparameter was provided in \cite{Kohler2008}.
\\ 

\noindent
\textbf{Random walk with restart (RWR). \cite{Kohler2008}} The random walk with restart is a discrete time random walk, where at every step there is a probability of returning to the initial state. The initial state is chosen to be a uniform distribution at the seed proteins, and the scores for proteins are their probability values at the steady-state distribution. 
In this model, the restart probability can be considered as a hyperparameter, however, we found that our results were not sensitive to it. We used a restart probability of $0.4$ for all experiments, as was done in \cite{Ghiassian2015diamond}.
\\

\noindent
\textbf{DIAMOnD (DIA). \cite{Ghiassian2015diamond}} DIAMOnD iteratively adds proteins to the disease module based on their connectivity significance to the seed proteins. Because DIAMOnD expands the module one protein at the time, it can rank proteins away from the immediate neighbourhood of the disease proteins. For our experiments, we used the extended version of the algorithm described in \cite{Ghiassian2015diamond}, weighting the seed proteins with a value of $\alpha=9$ (setting $\alpha \approx 10$ was recommended in the original paper).\\

\noindent
\textbf{Neighbourhood scoring (NEI). \cite{Navlakha2010}} In the neighbourhood method, each protein is assigned a score that is proportional to the percentage of its neighbours associated with the disease. Thus, this method is limited to scoring only the immediate neighbourhood of the seed proteins as all other proteins are given a score of zero. \\

\subsection{Metrics} \label{sec: metrics}

Since the ground truth of the disease modules is unknown, we proceeded to test the algorithms using cross-validation. For each disease, we randomly removed $50\%$ of the seed genes, and reserved these genes as positive test cases. The rest of the genes were used as negative testing data. In other words, after removing the $50\%$ of the seed genes, the non-seeds were ranked by sorting them in descending order according to their scores given by each method, and the genes with higher scores were deemed most likely to exist. This ranking was then compared to the evaluation set to see how well the positive test cases were ranked. This process was repeated 10 times for each disease, and the results were averaged (see below).

In order to compare the protein rankings of the methods under consideration, we used recall defined as 
\begin{align*}
\text{true positive rate = recall} = \dfrac{\text{TP}}{\text{TP} + \text{FN}},
\end{align*} 
where TP = true positive and FN = false negative. To calculate recall from the rankings, a threshold that serves as a cut-off rule has to be selected (the predictions above the thresholds are classified as positive and below it as negative). We consider thresholds up to $300$, which means that the top $300$ predictions are evaluated. The whole pipeline can be seen below in Figure \ref{fig:pipeline}.

There is, however, considerable variance in the recall values across different diseases, making the averaging of these values a less robust metric for measuring the performance of the methods across diseases. Therefore, to ensure a more comprehensive comparison of the methods across diseases, we calculated the mean reciprocal ranks, as was done in \cite{Agrawal2018}. This approach provides a more reliable way to evaluate the method's performance in a diverse range of scenarios.

For a method $k$ evaluated on a particular set of diseases $\mathcal{D}$, the mean reciprocal rank is defined by
\begin{equation*}
    \text{MMR}(k) = \frac{1}{|\mathcal{D}|} \sum_{d=1}^{|\mathcal{D}|} \frac{1}{R^k_d},
\end{equation*}
where $|\mathcal{D}|$ is the number of diseases in the set and $R^k_d$ is the rank of the $k$th method for disease $d$, relative to the average recalls (over 10 trials) of the other methods being considered. In this way, the number of diseases for which the highest recalls are achieved can be compared, rather than the raw recall values.

It is also worth noting that we do not use the area under the receiving-operator characteristics curve or area under the precision-recall curve as our main metrics for comparing different algorithms as is customary in most binary classification tasks. The reason for this constraint is that it is too costly for the DIAMOnD algorithm to assign scores to every protein in the network. 

\begin{figure*}[!htb] 
 \includegraphics[width=0.95\linewidth]{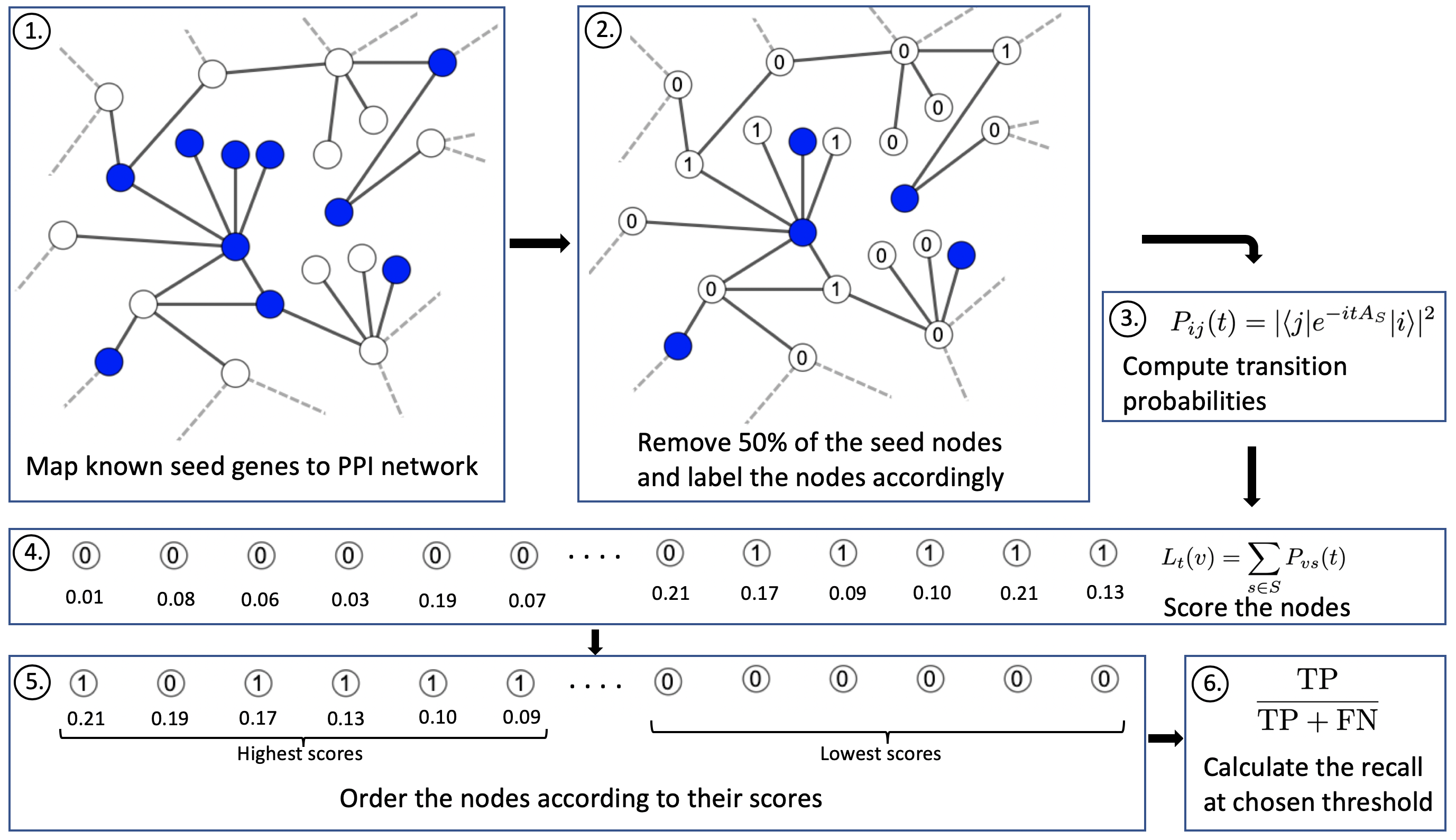}
 \caption{
 \textbf{Description of the algorithm and evaluation procedure}: 1) The seed genes are mapped to a given PPI network. 2) Half of the seed nodes are randomly selected and labelled as non-seeds. 3) Transition probabilities of the quantum walk are calculated for every pair of nodes in the network. 4) Genes are scored according to their seed node transition probabilities. 5) The genes are ranked from highest to lowest scores, with their ground truth labels preserved. 6) The recall value for a given threshold is calculated as the fraction of true seeds in the top $N$ scores (we use $N=25$ and $N=300$ in our results).}
 \label{fig:pipeline}
\end{figure*}

\section{Results} \label{Results}

Tables \ref{MRR25} and \ref{MRR300} present the average MRR values achieved over ten runs for each method across all considered disease sets and networks, specifically ranking the top 25 and top 300 nodes, respectively. Likewise, Tables \ref{Recall25} and \ref{Recall300} list the average Recall values for the respective numbers of ranked nodes of 25 and 300.

The QA model consistently outperforms other models across most disease sets and networks at both 25 and 300 ranked nodes, as indicated by the highest MRR values in most cases. The DIA model also demonstrates competitive performance in a few cases when 25 proteins are ranked. The RWR model's performance is generally lower than the QA model, but it excels in certain network and disease set combinations. On average, it also outperforms the DIA model. The performance of DK and NBR models is mixed, and they do not consistently outperform other models for any disease set and network combination. 

In the recall results in Tables \ref{Recall25} and \ref{Recall300}, the QA model consistently achieves the highest average recall values across most disease sets and networks when ranking 300 nodes. When ranking only 25 nodes, QA still outperforms other models in most cases, but struggles in the DGN disease set. Notably, DIA's performance remains most competitive in this scenario. While the DK and RWR models demonstrate good performance in a few cases, they lack consistent results overall. By contrast, the NBR model's performance remains consistently poor in all cases.

\begin{figure*}[!htb]
    \centering
    \includegraphics[width=0.95\linewidth]{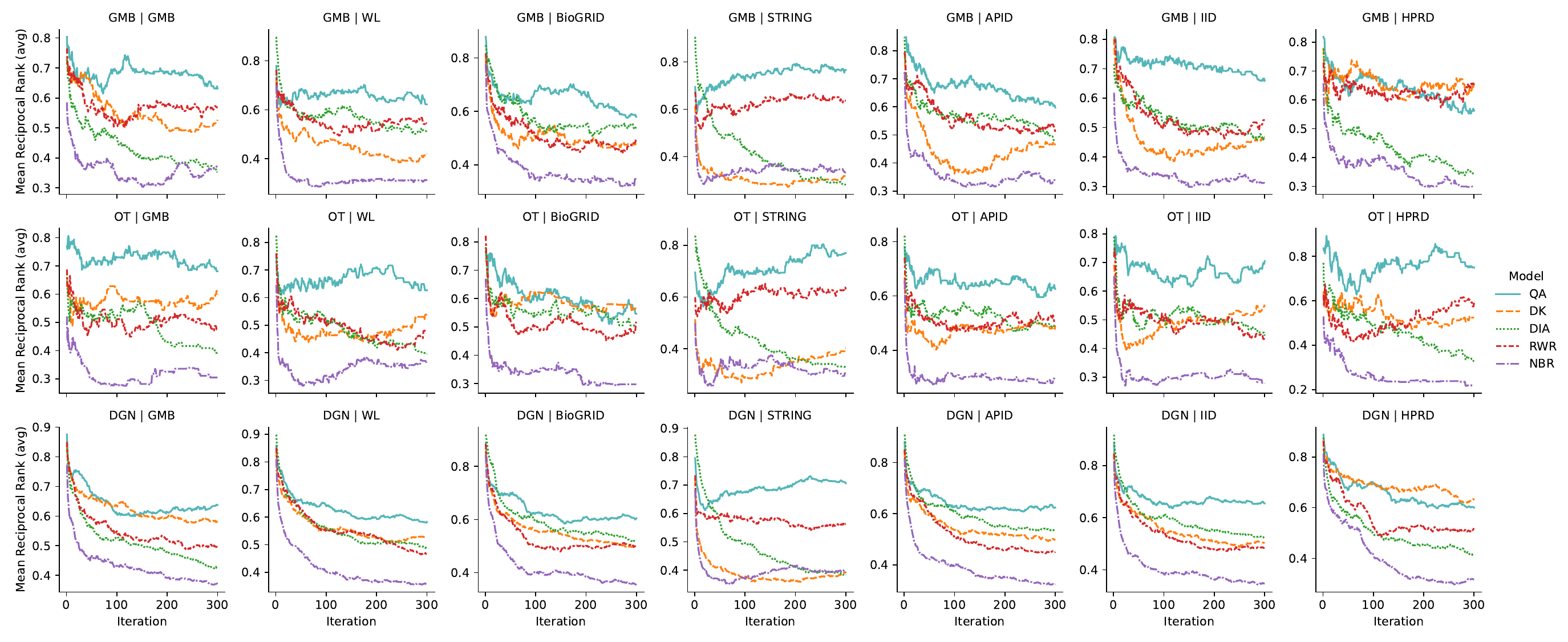}
    \caption{Mean reciprocal ranks averaged over 10 runs and all diseases for the three disease sets (rows) and seven networks (columns).}
    \label{mrrgrid}
\end{figure*}

\begin{figure*}[!htb]
    \centering
    \includegraphics[width=0.95\linewidth]{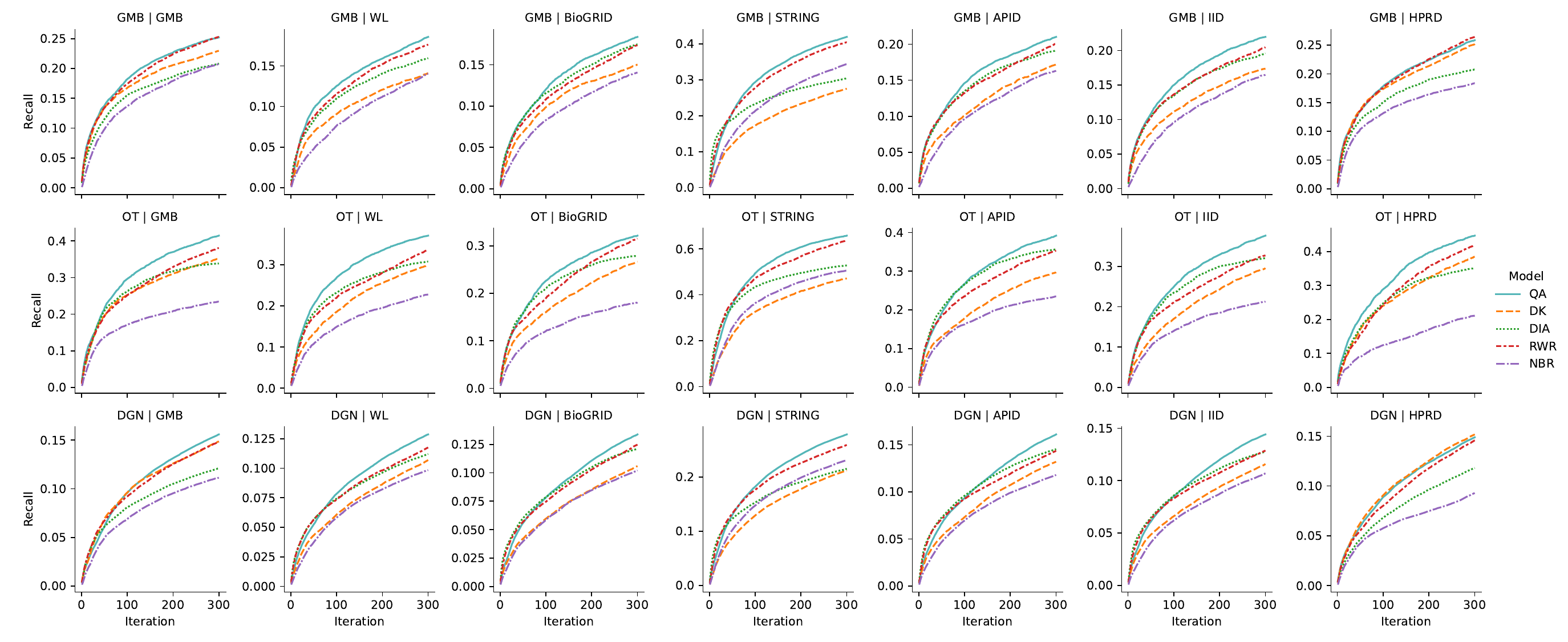}
    \caption{Recalls averaged over 10 runs and all diseases for the three disease sets (rows) and seven networks (columns).}
    \label{recallgrid}
\end{figure*}

Figures \ref{mrrgrid} and \ref{recallgrid} show a more global view of these results, plotting MRR and Recall as a function of the number of ranked nodes ranging from 1 to 300. Figure \ref{mrrgrid} illustrates the average MRR values obtained from the ten runs for all diseases in the three disease sets across the seven networks. Similarly, Figure \ref{recallgrid} displays the corresponding recall plots in the same experimental setup. The conclusion remains the same: The QA model very consistently outperforms other models when about 100 or more nodes are ranked and remains on par with other models when using less than 100 nodes.

In summary, the QA model outperforms other models in terms of both average Mean Reciprocal Rank (MRR) and average Recall across different disease sets and networks. The DIA, RWR, and DK also demonstrate good performance in some cases, but do not consistently outperform the QA model, especially when more than 100 nodes are ranked. Overall, the QA model appears to be the most robust and effective model for the given tasks.

\begin{table}[!htb]
\begin{tabular}{lllllll}
\toprule
   & Model &              QA &             DIA &     DK &    NBR &             RWR \\
Disease Set & Network &                 &                 &        &        &                 \\
\midrule
DGN & APID &           0.695 &  \textbf{0.700} &  0.666 &  0.488 &           0.677 \\
   & BioGRID &  \textbf{0.721} &           0.714 &  0.648 &  0.533 &           0.664 \\
   & HPRD &  \textbf{0.771} &           0.633 &  0.744 &  0.578 &           0.716 \\
   & GMB &  \textbf{0.756} &           0.650 &  0.696 &  0.500 &           0.641 \\
   & IID &  \textbf{0.705} &           0.693 &  0.655 &  0.489 &           0.666 \\
   & STRING &           0.618 &  \textbf{0.660} &  0.426 &  0.394 &           0.574 \\
   & WL &  \textbf{0.693} &           0.673 &  0.666 &  0.524 &           0.671 \\
GMB & APID &  \textbf{0.743} &           0.644 &  0.560 &  0.399 &           0.715 \\
   & BioGRID &  \textbf{0.682} &           0.653 &  0.545 &  0.468 &           0.599 \\
   & HPRD &  \textbf{0.709} &           0.555 &  0.667 &  0.450 &           0.635 \\
   & GMB &  \textbf{0.719} &           0.506 &  0.623 &  0.356 &           0.640 \\
   & IID &  \textbf{0.751} &           0.567 &  0.558 &  0.359 &           0.614 \\
   & STRING &  \textbf{0.662} &           0.573 &  0.333 &  0.332 &           0.585 \\
   & WL &           0.612 &           0.604 &  0.492 &  0.341 &  \textbf{0.637} \\
OT & APID &  \textbf{0.640} &           0.534 &  0.427 &  0.268 &           0.530 \\
   & BioGRID &           0.582 &  \textbf{0.612} &  0.541 &  0.328 &           0.571 \\
   & HPRD &  \textbf{0.774} &           0.491 &  0.583 &  0.332 &           0.446 \\
   & GMB &  \textbf{0.706} &           0.538 &  0.519 &  0.339 &           0.489 \\
   & IID &  \textbf{0.677} &           0.557 &  0.393 &  0.271 &           0.524 \\
   & STRING &           0.581 &  \textbf{0.646} &  0.322 &  0.257 &           0.527 \\
   & WL &  \textbf{0.623} &           0.529 &  0.468 &  0.329 &           0.526 \\ \hline
   & Average & \textbf{0.689} & 0.618 & 0.536 & 0.391 & 0.606 \\
\bottomrule
\end{tabular}
\caption{Mean reciprocal ranks for each model on each disease set and network, after $25$ nodes are scored.}
\label{MRR25}
\end{table}

\begin{table}[!htb]
\begin{tabular}{lllllll}
\toprule
   & Model &              QA &    DIA &              DK &    NBR &    RWR \\
Disease Set & Network &                 &        &                 &        &        \\
\midrule
DGN & APID &  \textbf{0.623} &  0.535 &           0.501 &  0.340 &  0.462 \\
   & BioGRID &  \textbf{0.606} &  0.533 &           0.506 &  0.371 &  0.510 \\
   & HPRD &           0.595 &  0.415 &  \textbf{0.640} &  0.322 &  0.526 \\
   & GMB &  \textbf{0.638} &  0.426 &           0.592 &  0.373 &  0.504 \\
   & IID &  \textbf{0.643} &  0.510 &           0.509 &  0.356 &  0.506 \\
   & STRING &  \textbf{0.712} &  0.390 &           0.389 &  0.397 &  0.558 \\
   & WL &  \textbf{0.573} &  0.491 &           0.547 &  0.360 &  0.475 \\
GMB & APID &  \textbf{0.626} &  0.510 &           0.432 &  0.338 &  0.499 \\
   & BioGRID &  \textbf{0.561} &  0.526 &           0.475 &  0.331 &  0.527 \\
   & HPRD &           0.531 &  0.368 &  \textbf{0.624} &  0.299 &  0.614 \\
   & GMB &  \textbf{0.631} &  0.381 &           0.500 &  0.378 &  0.591 \\
   & IID &  \textbf{0.617} &  0.435 &           0.481 &  0.329 &  0.522 \\
   & STRING &  \textbf{0.788} &  0.289 &           0.297 &  0.366 &  0.634 \\
   & WL &  \textbf{0.672} &  0.468 &           0.397 &  0.323 &  0.555 \\
OT & APID &  \textbf{0.611} &  0.491 &           0.474 &  0.290 &  0.510 \\
   & BioGRID &           0.515 &  0.503 &  \textbf{0.546} &  0.288 &  0.509 \\
   & HPRD &  \textbf{0.672} &  0.317 &           0.542 &  0.218 &  0.597 \\
   & GMB &  \textbf{0.622} &  0.420 &           0.547 &  0.312 &  0.514 \\
   & IID &  \textbf{0.643} &  0.418 &           0.539 &  0.277 &  0.470 \\
   & STRING &  \textbf{0.777} &  0.353 &           0.350 &  0.346 &  0.653 \\
   & WL &  \textbf{0.626} &  0.430 &           0.514 &  0.307 &  0.417 \\ \hline
   & Average & \textbf{0.625} & 0.456 & 0.491 & 0.330 & 0.529 \\
\bottomrule
\end{tabular}
\caption{Mean reciprocal ranks for each model on each disease set and network, after $300$ nodes are scored.}
\label{MRR300}
\end{table}

\begin{table}[!htb]
\begin{tabular}{lllllll}
\toprule
   & Model &              QA &             DIA &              DK &    NBR &             RWR \\
Disease Set & Network &                 &                 &                 &        &                 \\
\midrule
DGN & APID &           0.046 &  \textbf{0.057} &           0.041 &  0.030 &           0.056 \\
   & BioGRID &           0.038 &  \textbf{0.046} &           0.031 &  0.027 &           0.044 \\
   & HPRD &           0.044 &           0.034 &  \textbf{0.044} &  0.030 &           0.041 \\
   & GMB &           0.045 &           0.050 &           0.049 &  0.035 &  \textbf{0.051} \\
   & IID &           0.046 &  \textbf{0.055} &           0.042 &  0.028 &           0.054 \\
   & STRING &           0.094 &  \textbf{0.113} &           0.063 &  0.075 &           0.106 \\
   & WL &           0.038 &  \textbf{0.044} &           0.033 &  0.024 &           0.043 \\
GMB & APID &           0.074 &           0.079 &           0.057 &  0.038 &  \textbf{0.080} \\
   & BioGRID &  \textbf{0.066} &           0.061 &           0.049 &  0.032 &           0.061 \\
   & HPRD &  \textbf{0.111} &           0.101 &           0.110 &  0.083 &           0.110 \\
   & GMB &  \textbf{0.107} &           0.089 &           0.098 &  0.064 &           0.103 \\
   & IID &  \textbf{0.080} &           0.070 &           0.058 &  0.038 &           0.076 \\
   & STRING &           0.136 &  \textbf{0.157} &           0.078 &  0.087 &           0.148 \\
   & WL &  \textbf{0.072} &           0.063 &           0.051 &  0.036 &           0.071 \\
OT & APID &           0.135 &  \textbf{0.144} &           0.102 &  0.085 &           0.143 \\
   & BioGRID &           0.106 &  \textbf{0.119} &           0.086 &  0.061 &           0.111 \\
   & HPRD &  \textbf{0.137} &           0.113 &           0.103 &  0.061 &           0.100 \\
   & GMB &  \textbf{0.145} &           0.142 &           0.132 &  0.098 &           0.137 \\
   & IID &  \textbf{0.124} &           0.124 &           0.081 &  0.063 &           0.121 \\
   & STRING &           0.220 &  \textbf{0.265} &           0.134 &  0.138 &           0.261 \\
   & WL &  \textbf{0.139} &           0.129 &           0.097 &  0.072 &           0.123 \\ \hline
   & Average & 0.069 & \textbf{0.076} & 0.065 & 0.049 & 0.072 \\
\bottomrule
\end{tabular}
\caption{Average recall values when $25$ nodes are scored.}
\label{Recall25}
\end{table}

\begin{table}[!htb]
\begin{tabular}{lllllll}
\toprule
   & Model &              QA &    DIA &              DK &    NBR &             RWR \\
Disease Set & Network &                 &        &                 &        &                 \\
\midrule
DGN & APID &  \textbf{0.164} &  0.147 &           0.135 &  0.118 &           0.148 \\
   & BioGRID &  \textbf{0.133} &  0.120 &           0.105 &  0.101 &           0.125 \\
   & HPRD &           0.155 &  0.121 &  \textbf{0.160} &  0.095 &           0.154 \\
   & GMB &  \textbf{0.163} &  0.125 &           0.158 &  0.114 &           0.157 \\
   & IID &  \textbf{0.156} &  0.134 &           0.127 &  0.113 &           0.142 \\
   & STRING &  \textbf{0.299} &  0.235 &           0.224 &  0.252 &           0.278 \\
   & WL &  \textbf{0.137} &  0.116 &           0.119 &  0.101 &           0.125 \\
GMB & APID &  \textbf{0.226} &  0.209 &           0.178 &  0.173 &           0.214 \\
   & BioGRID &  \textbf{0.194} &  0.187 &           0.160 &  0.154 &           0.186 \\
   & HPRD &           0.268 &  0.223 &           0.260 &  0.195 &  \textbf{0.273} \\
   & GMB &           0.259 &  0.219 &           0.234 &  0.215 &  \textbf{0.261} \\
   & IID &  \textbf{0.227} &  0.193 &           0.182 &  0.171 &           0.212 \\
   & STRING &  \textbf{0.426} &  0.309 &           0.273 &  0.346 &           0.408 \\
   & WL &  \textbf{0.194} &  0.164 &           0.154 &  0.156 &           0.185 \\
OT & APID &  \textbf{0.400} &  0.355 &           0.303 &  0.236 &           0.360 \\
   & BioGRID &  \textbf{0.319} &  0.282 &           0.265 &  0.180 &           0.314 \\
   & HPRD &  \textbf{0.445} &  0.349 &           0.387 &  0.216 &           0.424 \\
   & GMB &  \textbf{0.416} &  0.342 &           0.353 &  0.241 &           0.388 \\
   & IID &  \textbf{0.373} &  0.319 &           0.291 &  0.216 &           0.333 \\
   & STRING &  \textbf{0.655} &  0.532 &           0.458 &  0.501 &           0.625 \\
   & WL &  \textbf{0.374} &  0.311 &           0.297 &  0.230 &           0.336 \\ \hline
   & Average & \textbf{0.285} & 0.239 & 0.216 & 0.181 & 0.257 \\
\bottomrule
\end{tabular}
\caption{Average recall values when $300$ nodes are scored.}
\label{Recall300}
\end{table}

\subsection{Coronary Artery Disease}
To further validate our methods, we used coronary artery disease (CAD) as a case study to demonstrate that the disease genes prioritized by the QA model are biologically relevant. We compiled a set of 81 seed genes for CAD derived from a meta-analysis of large-scale genome-wide association studies (GWAS) \cite{Nikpay2015}, mapped them to the WL human interactome (the largest of PPIs considered here), and used the QA model to prioritize disease genes for CAD. Of the 81 seed genes, 73 were be found in the WL human interactome. For this seed set, we optimized the parameters for the QA model using grid search and found that $t=0.11$ and $\alpha=5$ yield the best recall in cross-validation. Using these parameters we then considered the top 200 prioritized genes by QA. We also prioritized genes from the same starting seed set with the other models in this paper and found that of the top 200 genes prioritized by QA, 79 are not prioritized by any of the other methods. 

\begin{figure}[!ht]
    \centering
    \includegraphics[width=0.95\linewidth]{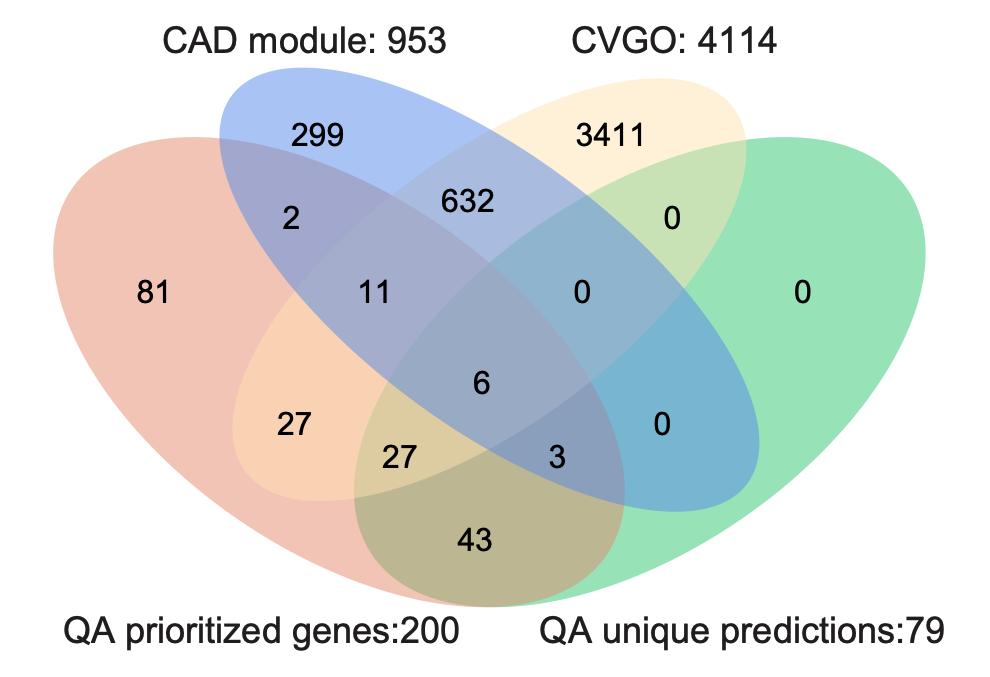}
    \caption{Overlap of the top 200 genes and 79 prioritized genes that were unique to QA with the CAD module compiled from OpenTargets, DisGeNet, and the genes from Cardiovascular Gene Ontology (CVGO) Annotation Initiative.}
    \label{fig: overlapCAD}
\end{figure}

We then examined the overlap of the top 200 genes and 79 prioritized genes that were unique to QA with the CAD module compiled from OpenTargets, DisGeNet, and the genes from Cardiovascular Gene Ontology (CVGO) Annotation Initiative (https://www.ebi.ac.uk/GOA/CVI). The top 200 prioritized genes and 79 uniquely predicted genes by the QA module have 22 and 9 overlapping genes with the CAD module, respectively (hypergeometric test, $p<0.001$ and $p<0.027$). These predictions are significantly enriched with genes from CVGO (hypergeometric test, $p<8.1 \cdot 10^{-5}$ and $p<2.4 \cdot 10^{-4}$ respectively for the top 200 predictions and 79 unique predictions) (Figure \ref{fig: overlapCAD}). This demonstrates the biological relevance of the prioritized genes for CAD. 

\section{Discussion}
\label{discussion}
\subsection{Ablation study on seed diagonals}

In Equation (\ref{eq:seed_diag}) we described our use of the hyperparameter $\alpha$, which has the effect of adding $\alpha$ self-edges to seed nodes before the quantum walks are performed. In this section, we explore the effect this hyperparameter has on the walk dynamics.

\begin{figure}[!htb]
    
    \centering
    \includegraphics[width=0.95\linewidth]{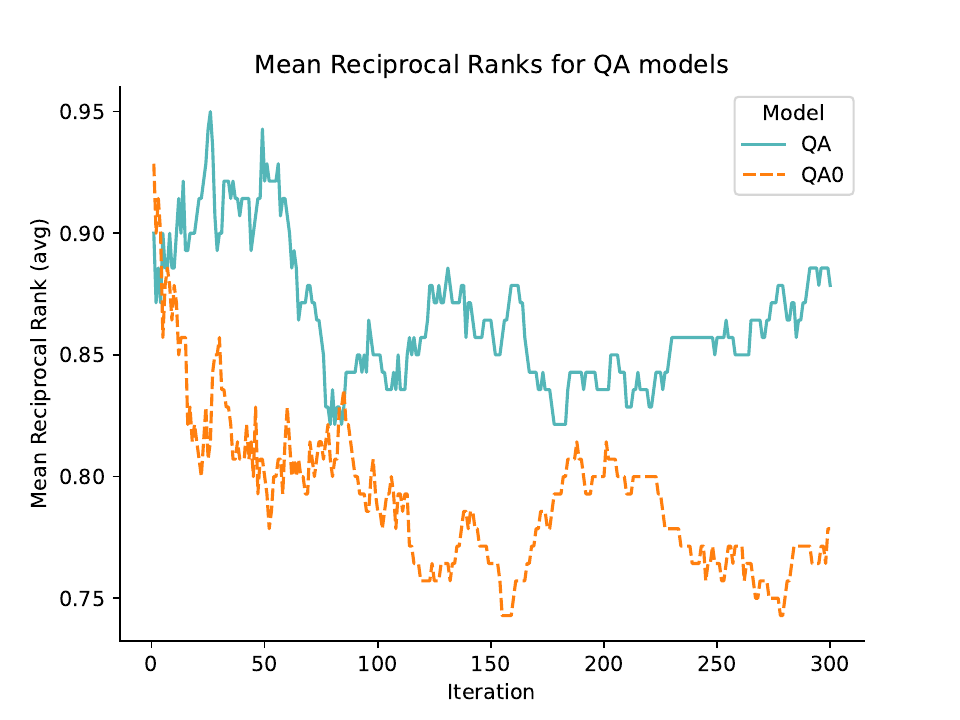}
    \caption{Mean reciprocal ranks averaged over 10 runs and all diseases in the GMB data set, using the GMB network.}
    \label{seed_diag_mrr}
\end{figure}

\begin{figure}[!htb]
    
    \centering
    \includegraphics[width=0.95\linewidth]{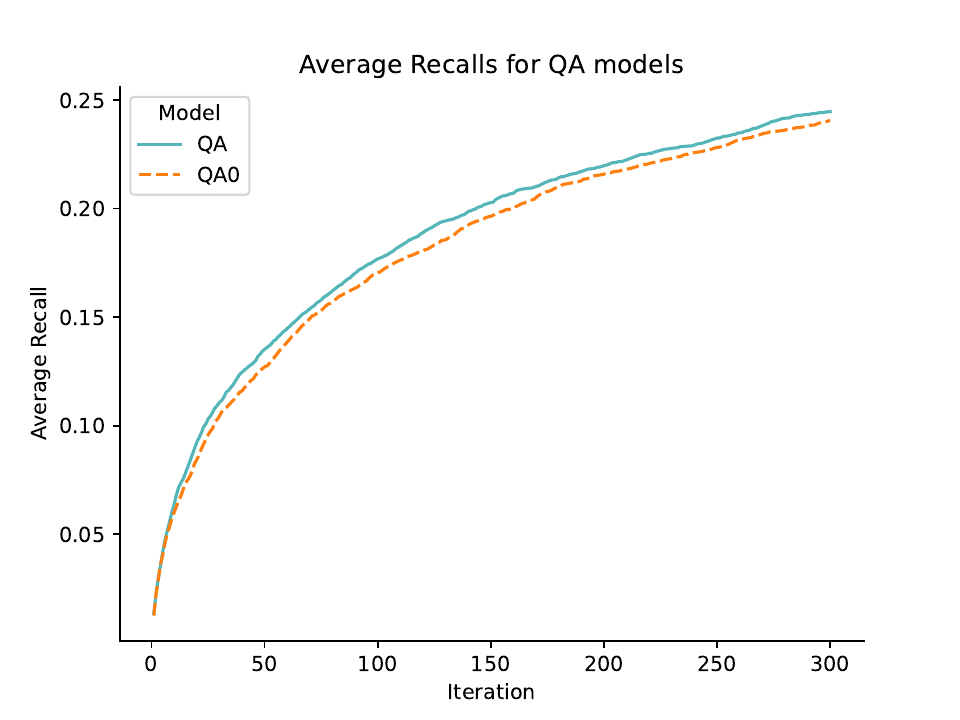}
    \caption{Recalls averaged over 10 runs and all diseases in the GMB data set, using the GMB network.}
    \label{seed_diag_recall}
\end{figure}

First, we demonstrate that our choice of $\alpha=5$ does, indeed, enhance the performance of our method by comparing the case used in our results $(\alpha=5)$ against the version of our model where seed diagonals are not treated in any particular way, i.e. $\alpha=0$. We compare these variations on the GMB disease set and network. The mean reciprocal ranks are shown in Figure \ref{seed_diag_mrr}, and their average recalls are compared in Figure \ref{seed_diag_recall}.

While Figures \ref{seed_diag_mrr} and \ref{seed_diag_recall} show that setting values to the seed diagonals does, indeed, improve overall performance, they do not offer any explanation as to why this should be the case. To this end, we offer the following hypothesis: a positive value of $\alpha$ allows the walker to remain more local for low-degree nodes. To justify this claim, we examine the mean distance travelled from seed node $s$ after time $t$, defined as 
\begin{equation}
\mu_s(t) = \sum_v d_{sv} P_{sv}(t),
\end{equation}
where $d_{sv}$ is the shortest path length from node $s$ to node $v$, and $P_{sv}(t)$ is the transition probability from $s$ to $v$ after time $t$, defined above in Equation (\ref{eq:QRWprobabilitymatrix}).

\begin{figure*}[!htb]
    \centering
    \includegraphics[width=0.95\linewidth]{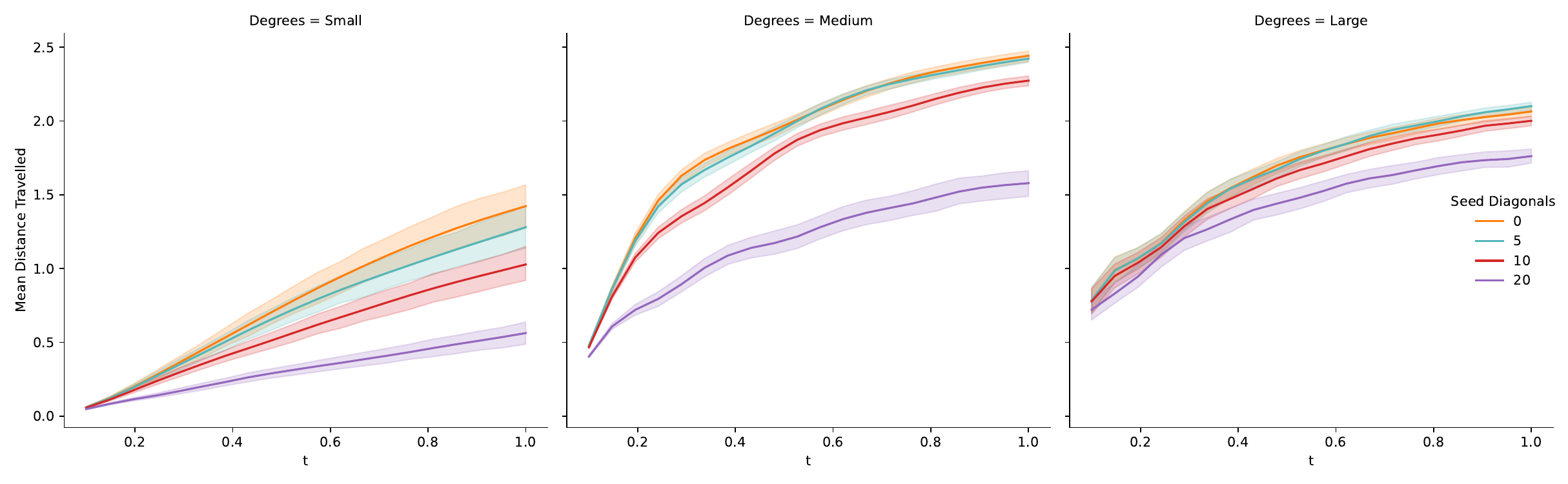}
    \caption{Mean distance travelled for various settings of the seed diagonals in the QA model for 3 different starting seeds: low (starting seed has degree at most 10); medium (starting seed has degree in $[50, 60]$); large (starting seed has degree in $[200, 300]$). In each of the 3 degree settings, curves are averaged over 50 starting seed runs.}
    \label{mdt_by_degrees}
\end{figure*}

In Figure \ref{mdt_by_degrees}, we show the results of the following experiment, conducted on the GMB PPI network: we choose a random low/medium/high degree node $s$ (defined as nodes with degrees in the range $[1, 10], [50,60], [200, 300]$, respectively) to use as a single starting seed node, then we compute the mean distance travelled from $s$ for multiple values of $t$ in the range $[0,1]$, and for four different settings of the hyperparameter $\alpha$. This process is repeated 50 times for each of the 3 degree ranges, and the results are averaged over the 50 runs, resulting in 4 curves for each of the $\alpha$ settings.

Two main conclusions can be drawn from Figure \ref{mdt_by_degrees}. Firstly, settings $\alpha=0$ and $\alpha=5$ have similar mean distance travelled curves for medium and high degree nodes, but not for low degree nodes, agreeing with our above hypothesis. Secondly, the highest value of $\alpha$ considered reduces the mean distance travelled in all cases. Thus, the setting of this hyperparameter should be chosen carefully, otherwise the walker may not have a chance to explore regions of the network not immediately adjacent to any seeds.

\begin{figure}[!htb]  
    \centering
    \includegraphics[width=0.95\linewidth]{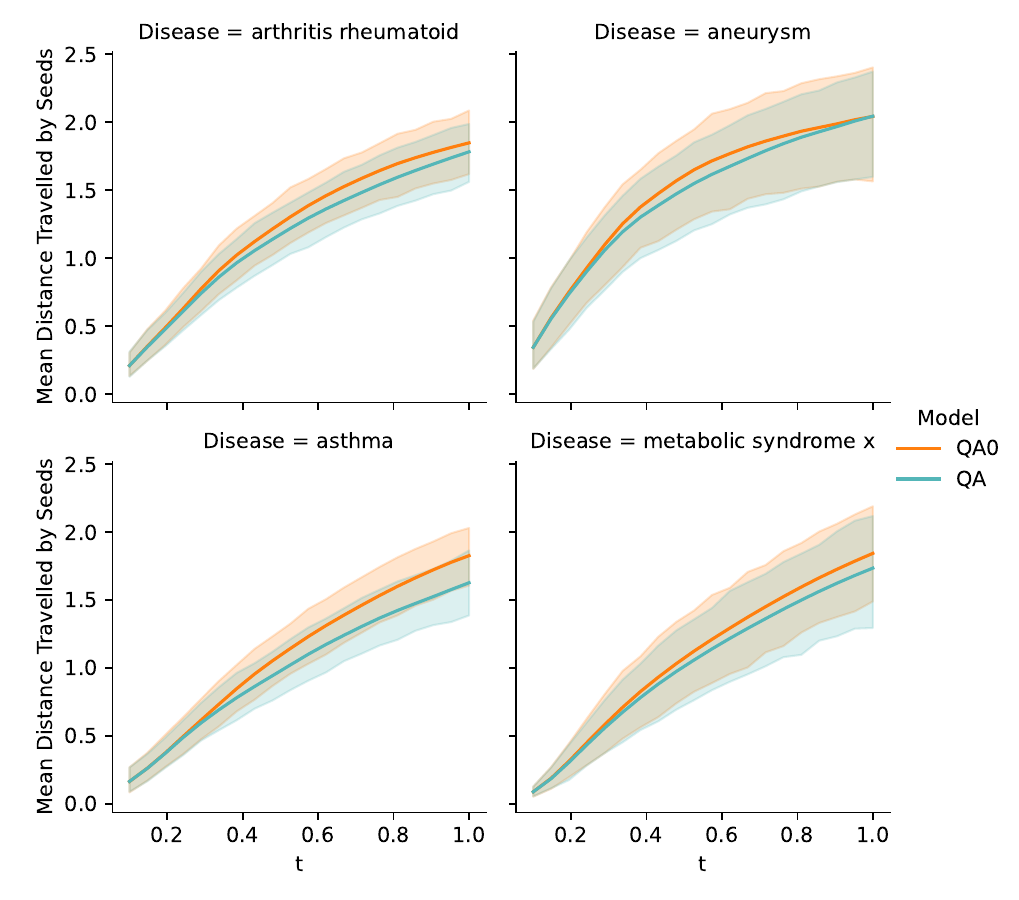}
    \caption{Mean distance travelled for 4 diseases from the GMB disease set on the GMB PPI network.}
    \label{mdt_4_d}
\end{figure}

The previous experiment considered the quantum walk dynamics for a single starting node. Next, we examined the more relevant case of having multiple seed nodes. More specifically, we choose four arbitrary diseases from the GMB disease set, and  we compute the mean distance travelled for each disease by averaging the resulting mean distance travelled curves over the seeds for each disease. In other words, for a disease with seed set $S$, we compute 

\begin{equation*}
    \frac{1}{|S|} \sum_{s \in S} \mu_s(t)
\end{equation*}
for several values of $t$. The results are shown in Figure \ref{mdt_4_d}. Indeed, we can see that, on average, quantum walkers will travel farther for $\alpha=0$ when compared to our setting of $\alpha=5.$

\section{Conclusion}

Our study introduces a novel algorithm for disease gene prioritization based on continuous-time quantum walks on PPI networks. The proposed algorithm demonstrates great performance compared to several well-known gene prioritization methods across multiple disease sets and various PPI networks. By encoding self-loops for the seed nodes into the underlying Hamiltonian, the quantum walker was shown to remain more local to the seed nodes, leading to improved performance.

The results indicate that the quantum walk-based algorithm can effectively prioritize disease genes by leveraging the structure of the PPI network and the known seed genes. The continuous-time quantum walk approach provides a flexible and efficient alternative to classical random walk methods more commonly used in various network biology tasks. However, further research and validation are necessary to fully understand the potential of quantum walks and their applicability to other biological network-related tasks.

Overall, the study contributes to the growing field of network medicine and computational methods for disease gene prioritization, highlighting the value of incorporating quantum-inspired algorithms in biological network analysis. With advances in quantum computing, future applications of quantum walks in this domain may hold even greater promise.

\section{Author Contributions}
M.G. and H.S. conceived of and implemented the algorithms. R.W. conducted the CAD enrichment analysis. M.G. and H.S. wrote the first version of the manuscript. All authors contributed to the scientific discussions and to the writing of the manuscript. All authors have read and agreed to the published version of the manuscript.
\bibliographystyle{acm}
\bibliography{bibs.bib}

\end{document}